\documentclass[%
reprint,
 superscriptaddress,
 amsmath,amssymb,
 aps,
 pra,
]{revtex4-1}

\usepackage{graphicx}
\usepackage{dcolumn}
\usepackage{bm}
\usepackage{color}
\usepackage[normalem]{ulem}

\begin{document}

\preprint{APS/123-QED}

\title{
Observation of Electric Octupole Emission Lines Strongly Enhanced\\
by an Anomalous Behavior of Cascading Contribution
}

\author{Hiroyuki A. Sakaue}
\affiliation{%
National Institute for Fusion Science, Toki, Gifu 509-5292, Japan
}%

\author{Daiji Kato}
\affiliation{%
National Institute for Fusion Science, Toki, Gifu 509-5292, Japan
}%
\affiliation{%
Department of Advanced Energy Engineering Science, Kyushu University, Fukuoka 816-8580, Japan
}%

\author{Izumi Murakami}
\affiliation{%
National Institute for Fusion Science, Toki, Gifu 509-5292, Japan
}%
\affiliation{%
Department of Fusion Science, The Graduate University for Advanced Studies (SOKENDAI),
Toki, Gifu 509-5292, Japan
}%

\author{Hayato Ohashi}
\affiliation{%
Institute of Liberal Arts and Sciences, University of Toyama, Toyama 
930-8555, Japan
}%

\author{Nobuyuki Nakamura}
\affiliation{%
Institute for Laser Science, The University of Electro-Communications, Tokyo 182-8585, Japan
}%

\date{\today}

\begin{abstract}
We present extreme ultraviolet spectra of Ag-like W$^{27+}$ observed with an electron beam ion trap.
In the spectra, the $4f_{7/2, 5/2}$ -- $5s$ transitions are identified as the first observation of spontaneous electric octupole emission.
Our theoretical investigation shows that the emission line intensity is strongly and specifically enhanced at the atomic number 74 by an anomalous behavior of cascading contribution to $5s$ via $5p\leftarrow5d$.
\end{abstract}

\maketitle


\section{\label{sec:introduction}Introduction}

Studies of electric dipole ($E1$) forbidden transitions are important not only for testing atomic physics theory describing the interaction of atoms or ions with multipole radiation fields, but also for several applications, such as plasma diagnostics \cite{Feldman1,Ralchenko3,Ralchenko4} and atomic clocks \cite{Margolis1,Takamoto1}.
The transition probability of forbidden transitions is generally too small for the decay to be observed in emission spectra of neutral atoms and low charged ions.
However, the probability increases rapidly with the atomic number $Z$ with a strong power law dependence.
As a typical example, the transition probability of the 1~$^1\!S_0$ -- 2~$^3\!S_1$ magnetic dipole ($M1$) transition in the He-like iso-electronic system has $Z^{10}$ dependence \cite{Beyer1}.
Thus observations of forbidden transitions have often been performed for highly charged ions to date.
For example, $M1$ transitions in highly charged heavy ions, such as iron and tungsten, are often used for the diagnostics of astrophysical and fusion plasmas \cite{DelZanna2,Kato8,Fujii1}.
Many $M1$ transitions have thus been observed and identified over a wide range of wavelengths so far \cite{Komatsu1,Ralchenko4}.
Electric and magnetic quadrupole ($E2$ and $M2$, respectively) transitions are also often observed in laboratory and astrophysical plasmas.
A typical example of $E2$ transition is $[3d^{10}]_{J=0}$ -- $[3d^9 4s]_{J=2}$ in Ni-like ions \cite{Klapisch1,Wyart2} whereas that of $M2$ transition is 1~$^1\!S_0$ -- 2~$^3\!P_2$ (the line often indicated as ``$x$'') in He-like ions \cite{Doschek1,Brown5}.

In contrast, observation of multipole transitions beyond quadrupole is quite limited even for highly charged ions.
There is only one example, which is the $[3d^{10}]_{J=0}$ -- $[3d^9 4s]_{J=3}$ magnetic octupole ($M3$) transition in Ni-like ions.
It was first observed in the x-ray region for Th$^{62+}$ ($Z$=90) and U$^{64+}$ (92) with an electron beam ion trap (EBIT) at Lawrence Livermore National Laboratory (LLNL) \cite{Beiersdorfer25}.
In an EBIT, trapped highly charged ions interact with a low-density (typically $10^{10}$ -- $10^{12}$ cm$^{-3}$) electron beam.
The collision frequency is typically in the order of 10 Hz, so that weak forbidden transitions with a transition probability down to $\sim 10$ s$^{-1}$ can be observed \cite{Urrutia7,Windberger2,Murata1}.
The observations with the LLNL EBIT showed that the $M3$ transition in Ni-like ions is indirectly excited through radiative cascades and can have an intensity comparable to $E1$ transitions depending on electron density.
The observation of the $M3$ transition was also conducted for Ni-like Xe$^{26+}$, Cs$^{27+}$, and Ba$^{28+}$ \cite{Trabert7}.
Time resolved measurements with a microcalorimeter mounted to the LLNL EBIT enabled acquisition of the decay lifetime of the metastable $[3d^9 4s]_{J=3}$ level.

Observations of an electric octupole ($E3$) transition in an atomic system have been reported for the $^2\!S_{1/2}$ -- $^2\!F_{7/2}$ transition in Yb$^+$ \cite{Roberts1,Roberts2,Hosaka2}.
In their observations, the transition was detected as an excitation driven with a laser.
The transition probability is so small ($\sim 10^{-9}$ s$^{-1}$ \cite{Roberts2}) that it is practically impossible to detect the emission.
There is no observation of spontaneous electric octupole ($E3$) emission so far even for highly charged ions.
In this paper, we present the first direct observation of spontaneous $E3$ emission lines performed for Ag-like W$^{27+}$.
In an emission spectrum of Ag-like W$^{27+}$ in the extreme ultraviolet (EUV) range observed with an EBIT, two lines are assigned to $4f_{7/2,5/2}$ -- $5s$ transitions, which can be realized by $E3$.
We also present the analysis based on collisional radiative (CR) model calculations.
The calculated result shows that an anomalous cascading contribution enhances the $E3$ emission intensity strongly and specifically at the atomic number $Z=74$, and hence enables us to observe the emission.

\section{\label{sec:exp}Experiments}
The present experiments were performed using a compact EBIT, called \mbox{CoBIT}　\cite{cobit}.
\mbox{CoBIT} mainly consists of an electron gun, an ion trap (drift tube), an electron collector, a superconducting coil, and a liquid nitrogen tank.
A high critical temperature superconducting Helmholtz-like coil, which can be used at the liquid nitrogen temperature, is mounted around the drift tube.
An electron beam emitted from the electron gun is accelerated (or decelerated) toward the drift tube while being compressed by a magnetic field produced by the superconducting coil.
After passing through the drift tube, the electron beam is collected by the electron collector.
In the present study, tungsten was injected into \mbox{CoBIT} as a sublimated vapor of tungsten hexacarbonyl W(CO)$_6$ through a gas injection system.

Emission from the trapped tungsten ions in the EUV region was observed with a flat-field grazing incidence grating spectrometer with a 1200 grooves/mm aberration-corrected concave grating (Hitachi 001-0660).
A back-illuminated charge coupled device (CCD) detector (Princeton Instruments PyLON:XO-2KB) was mounted at the focal position for detecting the diffracted EUV photons.
The CCD was cooled at -120~$^\circ$C by liquid nitrogen for suppressing the dark current.
Either aluminum or zirconium foil was placed in front of the grating for examining and removing the contribution of the second-order diffraction.
Furthermore, these metal filters cut the stray visible light from the cathode of the electron gun.
Wavelength calibration was carried out using well-known emission lines of highly charged Ar and O \cite{NISTdatabase2015}.
The uncertainty in the wavelength calibration was estimated to be less than $\pm$0.02 nm.

\section{\label{sec:crm}collisional radiative modeling}

To analyze the experimental spectra, we performed collisional radiative (CR) model calculations.
The line intensity of radiative transitions is expressed as the product of the transition probability and the fractional population of the upper level.
In the present CR model calculations, the dimension-less fractional population $n_i$ of the level $i$ was calculated by the following CR equilibrium equation,
\begin{equation}
\label{eq:inout}
n_i = \frac{n_e \sum_{j \neq i} C_{ij} n_j + \sum_{j > i} A_{ij} n_j}{n_e \sum_{j \neq i} C_{ji} + \sum_{j < i} A_{ji} }
\equiv \frac{C_{\rm in}+R_{\rm in}}{c_{\rm out}+r_{\rm out}} \equiv \frac{F_{\rm in}}{f_{\rm out}},
\end{equation}
where $C_{ij}$ and $A_{ij}$ are the electron impact (de)excitation rate coefficient and the radiative transition rate for the $i\leftarrow j$ transition, respectively, and $n_e$ electron density.
The electron collision rate coefficients were obtained by assuming the delta function distribution for the electron beam energy in \mbox{CoBIT}.
$F_{\rm in}$ stands for the inflow rate of populations to the level $i$, and consists of 
the collisional ($C_{\rm in}=n_e \sum_{j \neq i} C_{ij} n_j$) and radiative ($R_{\rm in}=\sum_{j > i} A_{ij} n_j$) inflow.
$f_{\rm out}$ stands for the rate of collisional ($c_{\rm out}=n_e \sum_{j \neq i} C_{ji}$) and radiative ($r_{\rm out}=\sum_{j < i} A_{ji}$) depopulations from the level $i$, which equals to the corresponding outflow rate divided by $n_i$.
Energy levels, radiative transition probabilities, and distorted-wave excitation and ionization cross sections were calculated with the Hebrew University Lawrence Livermore Atomic Code (HULLAC) \cite{HULLAC2}.
21,530 fine-structure levels of W$^{27+}$ below the ionization threshold (881.4 eV \cite{NIST}) were obtained by calculations with $4d^{10}4f$, $4d^{10}nl$ ($n=5$ -- 6, $l<n$), $4d^9 4f^2$, $4d^9 4f 5l$, $4d^9 5l^2$, $4d^8 4f^3$, and $4d^8 4f^2 5l$ configurations.


\section{\label{sec:results}Results and discussion}

\begin{figure}[tb]
\includegraphics[width=0.49\textwidth]{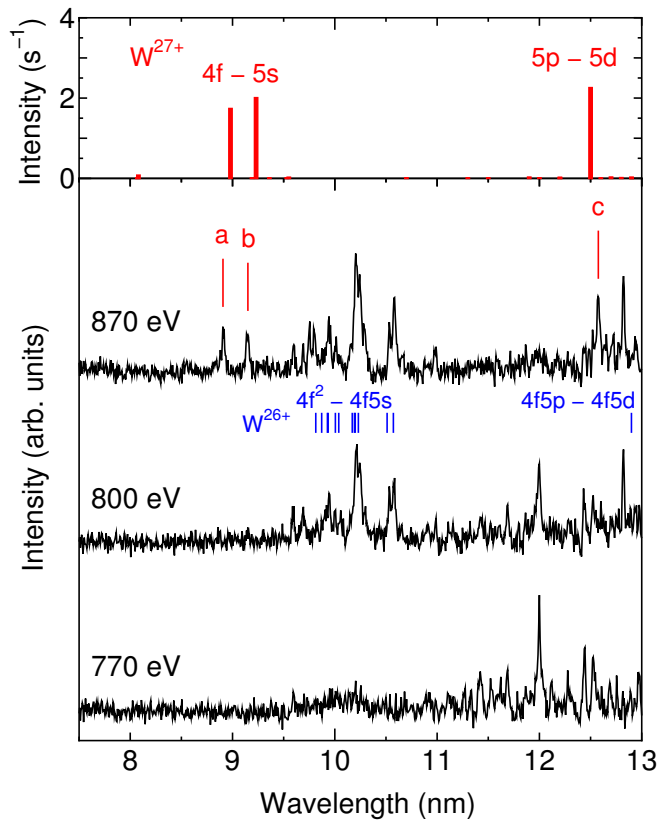}
\caption{\label{fig:spectra} EUV spectra of tungsten ions obtained with a compact electron beam ion trap with electron energies of 770, 800, and 870 eV.
The upper panel shows the CR model spectrum for W$^{27+}$.
The intensity in the upper panel is defined as the product of the $5s$ fractional population (dimensionless) and the $E3$ transition probability (s$^{-1}$).}
\end{figure}

Figure~\ref{fig:spectra} shows EUV spectra obtained at electron energies of 770, 800, and 870 eV.
The lines observed in the wavelength region 9.5 to 10.6 nm and a line at 12.8 nm in the 800eV spectrum were not observed at 770 eV.
Thus they should be assigned to W$^{26+}$ considering that the ionization potential of W$^{25+}$ is 784 eV \cite{NIST}.
Comparison with the calculated transition wavelengths shown as the blue vertical lines in the figure indicates that these lines should correspond to $4f^2$ -- $4f5s$ and $4f5p$ -- $4f5d$ transitions.
The $4f^2$ -- $4f5s$ transitions are strictly forbidden in a single configuration approach, but strongly enhanced by the configuration interaction between the $4f5s$ and $4f5d$ levels as discussed by Jonauskas {\it et al.} \cite{Jonauskas2}.
By increasing the electron energy further to 870 eV, three lines were additionally observed at around 8.9, 9.2, and 12.6 nm.
These lines should be assigned to W$^{27+}$ considering the ionization potential of W$^{26+}$ (833 eV) \cite{NIST}.
To identify these lines, the CR model spectrum is shown in the top panel of the figure.
As seen in the figure, in this wavelength region, the CR model calculation predicts three prominent lines, which should be assigned to the experimentally observed three lines.
One of them at around 12.6 nm is the $5p_{1/2}$ -- $5d_{3/2}$ $E1$ allowed transition, whereas the other two at 8.9~nm and 9.2~nm correspond to the $4f_{5/2, 7/2}$ -- $5s$ $E3$ transitions, respectively.
Table~\ref{tab:lines} lists the three lines in W$^{27+}$ observed in the present study with the present and available calculations \cite{Safronova7,Ding2}.
\begin{table*}
\caption{\label{tab:lines}Experimental and theoretical wavelengths of the emission lines in Ag-like W$^{27+}$ observed in the present study.
Calculated transition probabilities are also given in the last column.
}
\begin{ruledtabular}
\begin{tabular}{ccccccccc}
&&&&\multicolumn{4}{c}{Wavelength (nm)}&$A$ (s$^{-1}$)\\
\cline{5-8} \cline{9-9}
&\multicolumn{3}{c}{transition}&Exp.&\multicolumn{3}{c}{Theory}&Theory\\
\cline{2-4} \cline{5-5} \cline{6-8} \cline{9-9}
label&upper&lower&type&present&present&RMBPT\cite{Safronova7}&MCDF\cite{Ding2}&present\\
\hline
a&$5s$&$4f_{5/2}$&$E3$&8.91&8.982&8.905&9.040&$83$\\
&$5s$&$4f_{5/2}$&$M2$&-&-&-&-&$1.1\times10^{-3}$\\
b&$5s$&$4f_{7/2}$&$E3$&9.14&9.225&9.146&9.285&$96$\\
c&$5d_{3/2}$&$5p_{1/2}$&$E1$&12.59&12.533&12.589&--&$1.7\times10^{11}$\\
\end{tabular}
\end{ruledtabular}
\end{table*}
It is noted that the $4f_{5/2}$ -- $5s$ decay can be realized not only by $E3$ but also by $M2$.
However, as shown in the table, the transition probability by $M2$ ($1.1 \times 10^{-3}$~s$^{-1}$) is much smaller than that by $E3$ (83~s$^{-1}$), thus the transition is considered to be realized dominantly by $E3$.
As confirmed in the table, the wavelength values by Safronova \cite{Safronova7} show the best agreement with the present experimental values.

\begin{figure}[tb]
\includegraphics[width=0.45\textwidth]{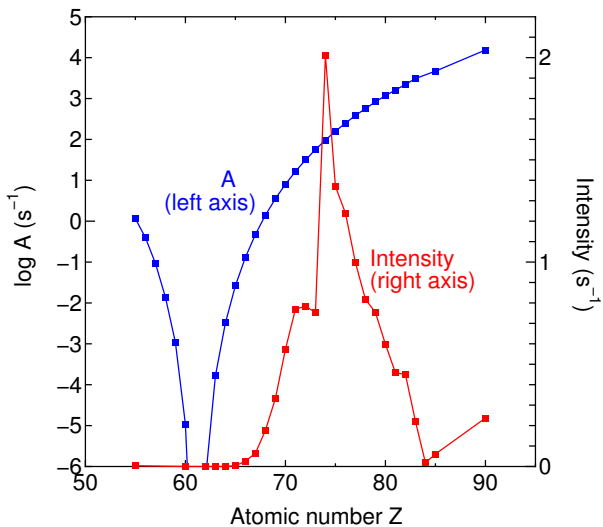}
\caption{\label{fig:zdep} Atomic number dependence of the transition probability (blue) and the intensity (red) of the $4f_{7/2}$ -- $5s$ $E3$ transition.
The definition of the intensity is the same as that in Fig.~\ref{fig:spectra}.
It is noted that the ground state is $5s$ and the upper state is $4f$ for $Z<62$,
whereas the ground state is $4f$ and the upper state is $5s$ for $Z\geqq62$.
}
\end{figure}

Although the transition probabilities of these $E3$ decays are much larger than that of, for example, the $^2\!F_{7/2}$ -- $^2\!S_{1/2}$ $E3$ transition in singly charged Yb$^+$ ($\sim 10^{-9}$~s$^{-1}$ \cite{Roberts1}) owing to their large transition energies, they are still much smaller than the transition probability of $E1$ allowed transitions.
However, the transition probability of forbidden transitions often has a strong $Z$ dependence, as described earlier.
The blue squares in Fig.~\ref{fig:zdep} show the atomic number dependence of the calculated transition probability of the $4f_{7/2}$ -- $5s$ $E3$ transition.
As expected, the transition probability increases quickly with $Z$.
The dependence on $Z$ is indeed about $Z^{40}$ at around $Z=74$.
This drastic dependence on $Z$ is considered to be caused by the fact that the level crossing between $4f$ and $5s$ exists at $Z\sim60$.
In the vicinity of the level crossing, the transition probability is almost zero as the transition energy is nearly zero.
When $Z$ is increased from the crossing, the energy interval $\Delta E$ between $4f$ and $5s$ rapidly increases.
This  results in the steep rise in the transition probability, which is approximately proportional to $\Delta E^5$ for this $E3$ transition.
Although the dependence becomes weaker as $Z$ increases, it is still $Z^{40}$ at $Z\sim74$.
One may thus expect that the intensity of the $E3$ lines should also increase when $Z$ is increased beyond 74.
Figure~\ref{fig:zdep} also plots the intensity of the $4f_{7/2}$ -- $5s$ $E3$ emission, calculated by the present CR model.
In each calculation, the electron energy was assumed to be just below the ionization energy of the Ag-like ion, and the electron density was fixed at $10^{10}$~cm$^{-3}$, which is the typical value in \mbox{CoBIT}.
The transition probability and the intensity of the $4f_{5/2}$ -- $5s$ $E3$ emission have almost the same $Z$-dependences although they are not shown in the figure.
As seen in the figure, in contrast to the expectation, the $4f_{7/2}$ -- $5s$ $E3$ emission line intensity has a sharp maximum at $Z=74$, and decreases quickly as $Z$ increases when $Z$ exceeds 74.

\begin{figure}[tb]
\includegraphics[width=0.45\textwidth]{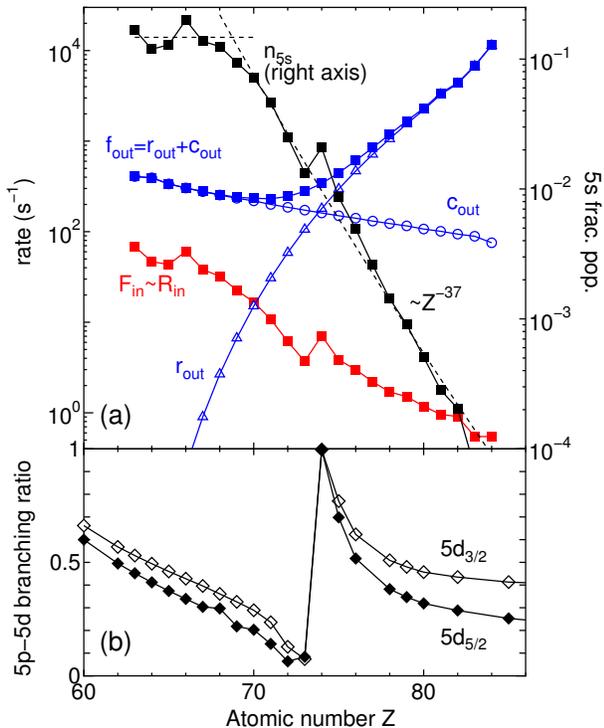}
\caption{\label{fig:inout}
(a) Inflow and depopulation rates for the metastable $5s$ level (left axis).
Fractional population of the $5s$ level determined by $F_{\rm in}/f_{\rm out}$ is also plotted (right axis).
See also Eq.~(\ref{eq:inout}).
(b) Branching ratios to $5p$ in the decay of the $5d_{3/2}$ (open diamond) and $5d_{5/2}$ (closed diamond) levels.
}
\end{figure}

In order to understand this strong $Z$-dependent behavior of the $E3$ intensity, population kinetics for $5s$ (the upper state of the $E3$ transition) is considered.
Direct collisional excitation rate from the ground state to $5s$ is negligibly small, and thus feeding by radiative cascades from upper states is required for populating $5s$.
The importance of radiative cascades was also confirmed for other multipole transitions, such as $E2$ \cite{Klapisch1} and $M3$ \cite{Beiersdorfer25} in low-density plasmas.
Fig.~\ref{fig:inout}~(a) shows theoretical values for inflow and depopulation rates (see Eq.~(\ref{eq:inout})) calculated for the $5s$ level.
The collisional inflow $C_{\rm in}$ is not shown because it is negligibly small (less than $10^{-1}$  s$^{-1}$) in this $Z$ region.
As shown in Eq.~(\ref{eq:inout}), the fractional population of the $5s$ level is determined by the ratio between the inflow rate $F_{\rm in}$ ($\sim R_{\rm in}$) and the depopulation rate $f_{\rm out}$.
The numerator $R_{\rm in}$ decreases as $Z$ increases for a whole range of $Z$ shown in Fig.~\ref{fig:inout}(a) because collisional excitation rates from the ground state to the higher levels decrease as $Z$ increases.
For $Z \leq 68$, the denominator $f_{\rm out}$ also decreases with a similar $Z$ dependence because $f_{\rm out}$ is dominated by the collisional depopulation $c_{\rm out}$, whose collisional excitation rates also decreases as $Z$ increases.
Thus, the fractional population of the $5s$ level, which is shown by the solid black squares in Fig.~\ref{fig:inout}~(a), is almost constant for $Z \leq 68$.
However, $f_{\rm out}$ starts to increase from $Z\sim 70$ due to the contribution of the radiative depopulation $r_{\rm out}$.
The increase of $f_{\rm out}$ shows the stronger $Z$ dependence than the decrease of $R_{\rm in}$.
As a result, the population is steeply decreased when $Z$ exceeds 70.
The resultant dependence is about $Z^{-37}$ as shown by the dashed line in the figure.
On the other hand, the $E3$ transition probability has a $Z^{a}$ dependence with $a\sim37$ at $Z\sim75$, whereas $a>37$ for $Z<75$ and $a<37$ for $Z>75$.
Thus, the $E3$ intensity, which is determined by the product of the $5s$ population and the transition probability, should have a maximum at $Z\sim75$.
However, such a maximum should be rather gentle, thus another mechanism should be needed to explain the sharp maximum at $Z=74$.

\begin{figure}[tb]
\includegraphics[width=0.5\textwidth]{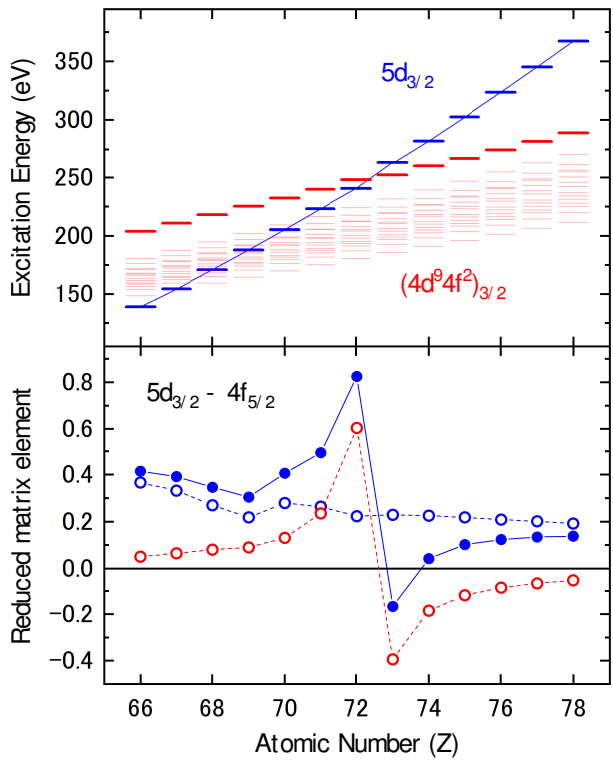}
\caption{\label{fig:lineS}
(a)	Energy level diagram of $5d_{3/2}$ (connected blue bars) and $\left(4d^9 4f^2\right)_{3/2}$ (red bars) states.
The $\left(4d^9 4f^2\right)_{3/2}$ state has 15 energy levels.
(b)	Reduced matrix elements of E1 transition between $5d_{3/2}$ and $4f_{5/2}$ states.
Open circles show $5d_{3/2}$ (blue) and $\left(4d^9 4f^2\right)_{3/2}$ (red) components, respectively.
The blue solid circles are the sum of the two components.
}
\end{figure}

Figure~\ref{fig:inout}(b) shows the branching ratios to $5p$ in the decay of the $5d$ levels.
As seen in the figure, they show anomalous $Z$ dependence with almost nought at $Z=73$ but almost unity at $Z=74$.
This behavior is due to an anomalous minimum at $Z=74$ in the $Z$-dependence of the $4f$ -- $5d$ transition probability.
The minimum can be understood as follows.
Figure~\ref{fig:lineS} (a) shows energy levels assigned to $5d_{3/2}$ and $\left( 4d^9 4f^2 \right)_{3/2}$ states as a function of $Z$.
The $5d_{3/2}$ level becomes quasi-degenerate with the highest level of the $\left(4d^9 4f^2\right)_{3/2}$ state in between $Z$ = 72 and 73, where the two levels have a strong configuration mixing.
Reduced matrix elements of the $E1$ transition between $5d_{3/2}$ and $4f_{5/2}$ levels exhibit a local irregularity at the level-crossing as shown in Fig.~\ref{fig:lineS} (b). The same irregularity has already been found in relativistic many-body perturbation calculations by Safronova {\it et al.} (see Fig. 2(c) of Ref. \cite{Safronova6}).
Here we investigate it in more detail to elucidate mechanisms of the anomalous minimum at $Z=74$. The transition matrix element can be decomposed into two components: the primary component of  $5d_{3/2}$ state and the complementary component which results from the configuration mixing of $\left( 4d^9 4f^2 \right)_{3/2}$ state.
The irregularity is caused by the $\left( 4d^9 4f^2 \right)_{3/2}$ component which increases resonantly and change the sign at the crossing.
The magnitude of the $\left( 4d^9 4f^2 \right)_{3/2}$ component decreases as Z becomes distant from the level-crossing, and becomes almost the same magnitude of the $5d_{3/2}$ component at $Z$ = 74. The two components of opposite signs cancel each other, akin to Cooper minima in photoionization cross sections, which results in the local minimum of the $4f_{5/2}$ -- $5d_{3/2}$ transition probability.
As a result of the minimum in the $4f$ -- $5d$ transition probability, almost all the $5d$ population decays to $5p$, and thus has a large contribution to the $5s$ population via $5p$ specifically at $Z=74$.
The enhancement in the $5s$ population, which can be confirmed as a deviation from the general trend ($Z^{-37}$ dependence shown by the dashed line in Fig.~\ref{fig:inout}(a)), is specifically large at $Z=74$, but rapidly decreases as $Z$ increases, and almost negligible at $Z\sim77$.
This behavior results in the sharp $Z$ dependence of the emission line intensity shown in Fig.~\ref{fig:zdep}.

\begin{figure}[tb]
\includegraphics[width=0.5\textwidth]{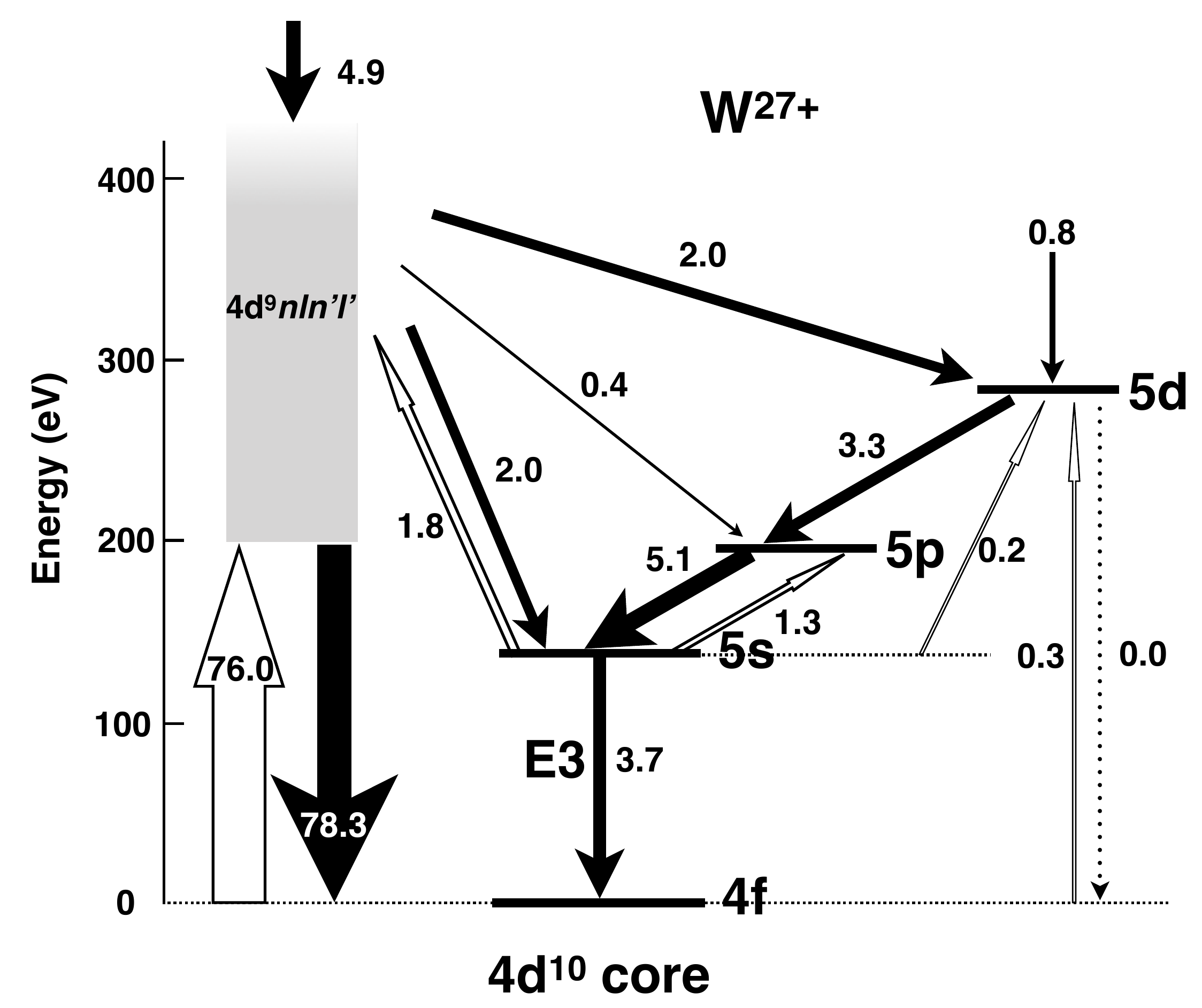}
\caption{\label{fig:flow_w}
Calculated energy levels and the population flows for Ag-like W$^{27+}$.
An electron energy of 870 eV and an density of $10^{10}$ cm$^{-3}$ are assumed in the CR model calculation.
The solid and open arrows represent radiative and collisional flows, respectively.
The number associated with each arrow represents the flow in s$^{-1}$.
}
\end{figure}

Figure~\ref{fig:flow_w} shows the energy levels of Ag-like W$^{27+}$ and the breakdown of the population flows calculated for an electron energy of 870 eV and a density of $10^{10}$ cm$^{-3}$.
The number associated with each arrow represents the flow in s$^{-1}$ defined by the product of the fractional population $n_i$ of the initial level $i$ and the rate of the transition to the levels $j$ ($n_e \sum_{j} C_{ji}$ for collisional transitions and $\sum_{j} A_{ji}$ for radiative transitions).
As understood from the figure, for the $4d^{10}5s$ level, which is the upper state of the $E3$ transitions, the collisional outflow to the $4d^{9} nl n'l'$ levels (1.8~s$^{-1}$) and the radiative inflow from the $4d^{9} nl n'l'$ levels (2.0~s$^{-1}$) are almost balanced.
The $E3$ flow is thus realized by the radiative inflow from the $4d^{10} 5p$ levels (5.1~s$^{-1}$), which exceeds the collisional outflow to the $4d^{10} 5p$ levels (1.3~s$^{-1}$).
The excess amount (3.8~s$^{-1}$) is predominantly due to the radiative inflow from the $4d^{10} 5d$ levels (3.3~s$^{-1}$).
If the anomalous minimum in the $4f$ -- $5d$ transition probability did not exist, the greater part of the radiative flow from the $4d^{10} 5d$ levels should fall into the $4d^{10} 4f$ ground state, and thus the prominent $E3$ emission could not be obtained.


\section{\label{sec:summary}Summary and outlook}

In summary, we have observed extreme ultraviolet spectra of Ag-like W$^{27+}$ with an electron beam ion trap, and identified the $4f_{7/2, 5/2}$ -- $5s$ transitions in the spectra as the first observation of spontaneous electric octupole ($E3$) emission lines.
Our collisional radiative model calculation has shown that the line intensity is strongly enhanced at $Z=74$ due to an anomalous behavior of the $4f$ -- $5d$ transition probability.
If it had not been for the anomalous behavior, we might have failed to observe the emission.

Tungsten is an important element for the spectroscopic diagnostics in the future ITER plasmaz \cite{Ralchenko4}.
The present $E3$ transitions should thus contribute to the future fusion plasma study.
For a fundamental atomic physics aspect, transition life time measurement for this $E3$ emission is desirable for understanding the interaction of ions with multipole radiation fields in more detail.

\section*{Acknowledgments}

This research was performed with the support and under the auspices of the NIFS Collaboration Research program (NIFS17KLPF057, NIFS17KLPF058, and NIFS17KBAF029), JSPS KAKENHI (Grant Number JP15K17728, JP16H04623, JP16H04028, and JP18H01201), and Fusion Engineering Research Project in NIFS (UFFF034).

\bibliography{ref}

\begin{thebibliography}{30}%
\makeatletter
\providecommand \@ifxundefined [1]{%
 \@ifx{#1\undefined}
}%
\providecommand \@ifnum [1]{%
 \ifnum #1\expandafter \@firstoftwo
 \else \expandafter \@secondoftwo
 \fi
}%
\providecommand \@ifx [1]{%
 \ifx #1\expandafter \@firstoftwo
 \else \expandafter \@secondoftwo
 \fi
}%
\providecommand \natexlab [1]{#1}%
\providecommand \enquote  [1]{``#1''}%
\providecommand \bibnamefont  [1]{#1}%
\providecommand \bibfnamefont [1]{#1}%
\providecommand \citenamefont [1]{#1}%
\providecommand \href@noop [0]{\@secondoftwo}%
\providecommand \href [0]{\begingroup \@sanitize@url \@href}%
\providecommand \@href[1]{\@@startlink{#1}\@@href}%
\providecommand \@@href[1]{\endgroup#1\@@endlink}%
\providecommand \@sanitize@url [0]{\catcode `\\12\catcode `\$12\catcode
  `\&12\catcode `\#12\catcode `\^12\catcode `\_12\catcode `\%12\relax}%
\providecommand \@@startlink[1]{}%
\providecommand \@@endlink[0]{}%
\providecommand \url  [0]{\begingroup\@sanitize@url \@url }%
\providecommand \@url [1]{\endgroup\@href {#1}{\urlprefix }}%
\providecommand \urlprefix  [0]{URL }%
\providecommand \Eprint [0]{\href }%
\providecommand \doibase [0]{http://dx.doi.org/}%
\providecommand \selectlanguage [0]{\@gobble}%
\providecommand \bibinfo  [0]{\@secondoftwo}%
\providecommand \bibfield  [0]{\@secondoftwo}%
\providecommand \translation [1]{[#1]}%
\providecommand \BibitemOpen [0]{}%
\providecommand \bibitemStop [0]{}%
\providecommand \bibitemNoStop [0]{.\EOS\space}%
\providecommand \EOS [0]{\spacefactor3000\relax}%
\providecommand \BibitemShut  [1]{\csname bibitem#1\endcsname}%
\let\auto@bib@innerbib\@empty
\bibitem [{\citenamefont {Feldman}\ \emph {et~al.}(1991)\citenamefont
  {Feldman}, \citenamefont {Indelicato},\ and\ \citenamefont
  {Sugar}}]{Feldman1}%
  \BibitemOpen
  \bibfield  {author} {\bibinfo {author} {\bibfnamefont {U.}~\bibnamefont
  {Feldman}}, \bibinfo {author} {\bibfnamefont {P.}~\bibnamefont {Indelicato}},
  \ and\ \bibinfo {author} {\bibfnamefont {J.}~\bibnamefont {Sugar}},\
  }\href@noop {} {\bibfield  {journal} {\bibinfo  {journal} {J. Opt. Soc. Am.
  B}\ }\textbf {\bibinfo {volume} {8}},\ \bibinfo {pages} {3} (\bibinfo {year}
  {1991})}\BibitemShut {NoStop}%
\bibitem [{\citenamefont {Ralchenko}(2007)}]{Ralchenko3}%
  \BibitemOpen
  \bibfield  {author} {\bibinfo {author} {\bibfnamefont {Y.}~\bibnamefont
  {Ralchenko}},\ }\href@noop {} {\bibfield  {journal} {\bibinfo  {journal}
  {Journal of Physics B: Atomic, Molecular and Optical Physics}\ }\textbf
  {\bibinfo {volume} {40}},\ \bibinfo {pages} {F175} (\bibinfo {year}
  {2007})}\BibitemShut {NoStop}%
\bibitem [{\citenamefont {Ralchenko}\ \emph {et~al.}(2011)\citenamefont
  {Ralchenko}, \citenamefont {Dragani\ifmmode~\acute{c}\else \'{c}\fi{}},
  \citenamefont {Osin}, \citenamefont {Gillaspy},\ and\ \citenamefont
  {Reader}}]{Ralchenko4}%
  \BibitemOpen
  \bibfield  {author} {\bibinfo {author} {\bibfnamefont {Y.}~\bibnamefont
  {Ralchenko}}, \bibinfo {author} {\bibfnamefont {I.~N.}\ \bibnamefont
  {Dragani\ifmmode~\acute{c}\else \'{c}\fi{}}}, \bibinfo {author}
  {\bibfnamefont {D.}~\bibnamefont {Osin}}, \bibinfo {author} {\bibfnamefont
  {J.~D.}\ \bibnamefont {Gillaspy}}, \ and\ \bibinfo {author} {\bibfnamefont
  {J.}~\bibnamefont {Reader}},\ }\href@noop {} {\bibfield  {journal} {\bibinfo
  {journal} {Phys. Rev. A}\ }\textbf {\bibinfo {volume} {83}},\ \bibinfo
  {pages} {032517} (\bibinfo {year} {2011})}\BibitemShut {NoStop}%
\bibitem [{\citenamefont {Margolis}(2009)}]{Margolis1}%
  \BibitemOpen
  \bibfield  {author} {\bibinfo {author} {\bibfnamefont {H.~S.}\ \bibnamefont
  {Margolis}},\ }\href@noop {} {\bibfield  {journal} {\bibinfo  {journal}
  {Journal of Physics B: Atomic, Molecular and Optical Physics}\ }\textbf
  {\bibinfo {volume} {42}},\ \bibinfo {pages} {154017} (\bibinfo {year}
  {2009})}\BibitemShut {NoStop}%
\bibitem [{\citenamefont {Takamoto}\ \emph {et~al.}(2005)\citenamefont
  {Takamoto}, \citenamefont {Hong}, \citenamefont {Higashi},\ and\
  \citenamefont {Katori}}]{Takamoto1}%
  \BibitemOpen
  \bibfield  {author} {\bibinfo {author} {\bibfnamefont {M.}~\bibnamefont
  {Takamoto}}, \bibinfo {author} {\bibfnamefont {F.-L.}\ \bibnamefont {Hong}},
  \bibinfo {author} {\bibfnamefont {R.}~\bibnamefont {Higashi}}, \ and\
  \bibinfo {author} {\bibfnamefont {H.}~\bibnamefont {Katori}},\ }\href@noop {}
  {\bibfield  {journal} {\bibinfo  {journal} {Nature}\ }\textbf {\bibinfo
  {volume} {435}},\ \bibinfo {pages} {321} (\bibinfo {year}
  {2005})}\BibitemShut {NoStop}%
\bibitem [{\citenamefont {Beyer}\ \emph {et~al.}(1997)\citenamefont {Beyer},
  \citenamefont {Kluge},\ and\ \citenamefont {Shevelko}}]{Beyer1}%
  \BibitemOpen
  \bibfield  {author} {\bibinfo {author} {\bibfnamefont {H.~F.}\ \bibnamefont
  {Beyer}}, \bibinfo {author} {\bibfnamefont {H.~J.}\ \bibnamefont {Kluge}}, \
  and\ \bibinfo {author} {\bibfnamefont {V.~P.}\ \bibnamefont {Shevelko}},\
  }\href@noop {} {\emph {\bibinfo {title} {X-ray Radiation of Highly Charged
  Ions}}}\ (\bibinfo  {publisher} {Springer},\ \bibinfo {address} {Berlin},\
  \bibinfo {year} {1997})\BibitemShut {NoStop}%
\bibitem [{\citenamefont {{Del~Zanna}}\ and\ \citenamefont
  {DeLuca}(2018)}]{DelZanna2}%
  \BibitemOpen
  \bibfield  {author} {\bibinfo {author} {\bibfnamefont {G.}~\bibnamefont
  {{Del~Zanna}}}\ and\ \bibinfo {author} {\bibfnamefont {E.~E.}\ \bibnamefont
  {DeLuca}},\ }\href@noop {} {\bibfield  {journal} {\bibinfo  {journal} {The
  Astrophysical Journal}\ }\textbf {\bibinfo {volume} {852}},\ \bibinfo {pages}
  {52} (\bibinfo {year} {2018})}\BibitemShut {NoStop}%
\bibitem [{\citenamefont {Kato}\ \emph {et~al.}(2013)\citenamefont {Kato},
  \citenamefont {Goto}, \citenamefont {Morita}, \citenamefont {Murakami},
  \citenamefont {Sakaue}, \citenamefont {Ding}, \citenamefont {Sudo},
  \citenamefont {Suzuki}, \citenamefont {Tamura}, \citenamefont {Nakamura},
  \citenamefont {Watanabe},\ and\ \citenamefont {Koike}}]{Kato8}%
  \BibitemOpen
  \bibfield  {author} {\bibinfo {author} {\bibfnamefont {D.}~\bibnamefont
  {Kato}}, \bibinfo {author} {\bibfnamefont {M.}~\bibnamefont {Goto}}, \bibinfo
  {author} {\bibfnamefont {S.}~\bibnamefont {Morita}}, \bibinfo {author}
  {\bibfnamefont {I.}~\bibnamefont {Murakami}}, \bibinfo {author}
  {\bibfnamefont {H.~A.}\ \bibnamefont {Sakaue}}, \bibinfo {author}
  {\bibfnamefont {X.~B.}\ \bibnamefont {Ding}}, \bibinfo {author}
  {\bibfnamefont {S.}~\bibnamefont {Sudo}}, \bibinfo {author} {\bibfnamefont
  {C.}~\bibnamefont {Suzuki}}, \bibinfo {author} {\bibfnamefont
  {N.}~\bibnamefont {Tamura}}, \bibinfo {author} {\bibfnamefont
  {N.}~\bibnamefont {Nakamura}}, \bibinfo {author} {\bibfnamefont
  {H.}~\bibnamefont {Watanabe}}, \ and\ \bibinfo {author} {\bibfnamefont
  {F.}~\bibnamefont {Koike}},\ }\href@noop {} {\bibfield  {journal} {\bibinfo
  {journal} {Physica Scripta}\ }\textbf {\bibinfo {volume} {2013}},\ \bibinfo
  {pages} {014081} (\bibinfo {year} {2013})}\BibitemShut {NoStop}%
\bibitem [{\citenamefont {Fujii}\ \emph {et~al.}(2015)\citenamefont {Fujii},
  \citenamefont {Takahashi}, \citenamefont {Nakai}, \citenamefont {Kato},
  \citenamefont {Goto}, \citenamefont {Morita}, \citenamefont {Hasuo},\ and\
  \citenamefont {{LHD Experiment Group}}}]{Fujii1}%
  \BibitemOpen
  \bibfield  {author} {\bibinfo {author} {\bibfnamefont {K.}~\bibnamefont
  {Fujii}}, \bibinfo {author} {\bibfnamefont {Y.}~\bibnamefont {Takahashi}},
  \bibinfo {author} {\bibfnamefont {Y.}~\bibnamefont {Nakai}}, \bibinfo
  {author} {\bibfnamefont {D.}~\bibnamefont {Kato}}, \bibinfo {author}
  {\bibfnamefont {M.}~\bibnamefont {Goto}}, \bibinfo {author} {\bibfnamefont
  {S.}~\bibnamefont {Morita}}, \bibinfo {author} {\bibfnamefont
  {M.}~\bibnamefont {Hasuo}}, \ and\ \bibinfo {author} {\bibnamefont {{LHD
  Experiment Group}}},\ }\href@noop {} {\bibfield  {journal} {\bibinfo
  {journal} {Physica Scripta}\ }\textbf {\bibinfo {volume} {90}},\ \bibinfo
  {pages} {125403} (\bibinfo {year} {2015})}\BibitemShut {NoStop}%
\bibitem [{\citenamefont {Komatsu}\ \emph {et~al.}(2011)\citenamefont
  {Komatsu}, \citenamefont {Sakoda}, \citenamefont {Nakamura}, \citenamefont
  {Sakaue}, \citenamefont {Din}, \citenamefont {Kato}, \citenamefont
  {Murakami},\ and\ \citenamefont {Koike}}]{Komatsu1}%
  \BibitemOpen
  \bibfield  {author} {\bibinfo {author} {\bibfnamefont {A.}~\bibnamefont
  {Komatsu}}, \bibinfo {author} {\bibfnamefont {J.}~\bibnamefont {Sakoda}},
  \bibinfo {author} {\bibfnamefont {N.}~\bibnamefont {Nakamura}}, \bibinfo
  {author} {\bibfnamefont {H.~A.}\ \bibnamefont {Sakaue}}, \bibinfo {author}
  {\bibfnamefont {X.-B.}\ \bibnamefont {Din}}, \bibinfo {author} {\bibfnamefont
  {D.}~\bibnamefont {Kato}}, \bibinfo {author} {\bibfnamefont {I.}~\bibnamefont
  {Murakami}}, \ and\ \bibinfo {author} {\bibfnamefont {F.}~\bibnamefont
  {Koike}},\ }\href@noop {} {\bibfield  {journal} {\bibinfo  {journal} {Phys.
  Scr.}\ }\textbf {\bibinfo {volume} {T144}},\ \bibinfo {pages} {014012}
  (\bibinfo {year} {2011})}\BibitemShut {NoStop}%
\bibitem [{\citenamefont {Klapisch}\ \emph {et~al.}(1978)\citenamefont
  {Klapisch}, \citenamefont {Schwob}, \citenamefont {Finkenthal}, \citenamefont
  {Fraenkel}, \citenamefont {Egert}, \citenamefont {Bar-Shalom}, \citenamefont
  {Breton}, \citenamefont {DeMichelis},\ and\ \citenamefont
  {Mattioli}}]{Klapisch1}%
  \BibitemOpen
  \bibfield  {author} {\bibinfo {author} {\bibfnamefont {M.}~\bibnamefont
  {Klapisch}}, \bibinfo {author} {\bibfnamefont {J.~L.}\ \bibnamefont
  {Schwob}}, \bibinfo {author} {\bibfnamefont {M.}~\bibnamefont {Finkenthal}},
  \bibinfo {author} {\bibfnamefont {B.~S.}\ \bibnamefont {Fraenkel}}, \bibinfo
  {author} {\bibfnamefont {S.}~\bibnamefont {Egert}}, \bibinfo {author}
  {\bibfnamefont {A.}~\bibnamefont {Bar-Shalom}}, \bibinfo {author}
  {\bibfnamefont {C.}~\bibnamefont {Breton}}, \bibinfo {author} {\bibfnamefont
  {C.}~\bibnamefont {DeMichelis}}, \ and\ \bibinfo {author} {\bibfnamefont
  {M.}~\bibnamefont {Mattioli}},\ }\href@noop {} {\bibfield  {journal}
  {\bibinfo  {journal} {Phys. Rev. Lett.}\ }\textbf {\bibinfo {volume} {41}},\
  \bibinfo {pages} {403} (\bibinfo {year} {1978})}\BibitemShut {NoStop}%
\bibitem [{\citenamefont {Wyart}\ \emph {et~al.}(1986)\citenamefont {Wyart},
  \citenamefont {Bauche-Arnoult}, \citenamefont {Gauthier}, \citenamefont
  {Geindre}, \citenamefont {Monier}, \citenamefont {Klapisch}, \citenamefont
  {Bar-Shalom},\ and\ \citenamefont {Cohn}}]{Wyart2}%
  \BibitemOpen
  \bibfield  {author} {\bibinfo {author} {\bibfnamefont {J.-F.}\ \bibnamefont
  {Wyart}}, \bibinfo {author} {\bibfnamefont {C.}~\bibnamefont
  {Bauche-Arnoult}}, \bibinfo {author} {\bibfnamefont {J.-C.}\ \bibnamefont
  {Gauthier}}, \bibinfo {author} {\bibfnamefont {J.-P.}\ \bibnamefont
  {Geindre}}, \bibinfo {author} {\bibfnamefont {P.}~\bibnamefont {Monier}},
  \bibinfo {author} {\bibfnamefont {M.}~\bibnamefont {Klapisch}}, \bibinfo
  {author} {\bibfnamefont {A.}~\bibnamefont {Bar-Shalom}}, \ and\ \bibinfo
  {author} {\bibfnamefont {A.}~\bibnamefont {Cohn}},\ }\href@noop {} {\bibfield
   {journal} {\bibinfo  {journal} {Phys. Rev. A}\ }\textbf {\bibinfo {volume}
  {34}},\ \bibinfo {pages} {701} (\bibinfo {year} {1986})}\BibitemShut
  {NoStop}%
\bibitem [{\citenamefont {Doschek}\ and\ \citenamefont
  {Tanaka}(1987)}]{Doschek1}%
  \BibitemOpen
  \bibfield  {author} {\bibinfo {author} {\bibfnamefont {G.~A.}\ \bibnamefont
  {Doschek}}\ and\ \bibinfo {author} {\bibfnamefont {K.}~\bibnamefont
  {Tanaka}},\ }\href@noop {} {\bibfield  {journal} {\bibinfo  {journal} {The
  Astrophysical Journal}\ }\textbf {\bibinfo {volume} {323}},\ \bibinfo {pages}
  {799} (\bibinfo {year} {1987})}\BibitemShut {NoStop}%
\bibitem [{\citenamefont {Brown}\ \emph {et~al.}(1989)\citenamefont {Brown},
  \citenamefont {Feldman}, \citenamefont {Doschek}, \citenamefont {Seely},
  \citenamefont {LaVilla}, \citenamefont {Jacobs}, \citenamefont {Henderson},
  \citenamefont {Knapp}, \citenamefont {Marrs}, \citenamefont {Beiersdorfer},\
  and\ \citenamefont {Levine}}]{Brown5}%
  \BibitemOpen
  \bibfield  {author} {\bibinfo {author} {\bibfnamefont {C.~M.}\ \bibnamefont
  {Brown}}, \bibinfo {author} {\bibfnamefont {U.}~\bibnamefont {Feldman}},
  \bibinfo {author} {\bibfnamefont {G.~A.}\ \bibnamefont {Doschek}}, \bibinfo
  {author} {\bibfnamefont {J.~F.}\ \bibnamefont {Seely}}, \bibinfo {author}
  {\bibfnamefont {R.~E.}\ \bibnamefont {LaVilla}}, \bibinfo {author}
  {\bibfnamefont {V.~L.}\ \bibnamefont {Jacobs}}, \bibinfo {author}
  {\bibfnamefont {J.~R.}\ \bibnamefont {Henderson}}, \bibinfo {author}
  {\bibfnamefont {D.~A.}\ \bibnamefont {Knapp}}, \bibinfo {author}
  {\bibfnamefont {R.~E.}\ \bibnamefont {Marrs}}, \bibinfo {author}
  {\bibfnamefont {P.}~\bibnamefont {Beiersdorfer}}, \ and\ \bibinfo {author}
  {\bibfnamefont {M.~A.}\ \bibnamefont {Levine}},\ }\href@noop {} {\bibfield
  {journal} {\bibinfo  {journal} {Phys. Rev. A}\ }\textbf {\bibinfo {volume}
  {40}},\ \bibinfo {pages} {4089} (\bibinfo {year} {1989})}\BibitemShut
  {NoStop}%
\bibitem [{\citenamefont {Beiersdorfer}\ \emph {et~al.}(1991)\citenamefont
  {Beiersdorfer}, \citenamefont {Osterheld}, \citenamefont {Scofield},
  \citenamefont {Wargelin},\ and\ \citenamefont {Marrs}}]{Beiersdorfer25}%
  \BibitemOpen
  \bibfield  {author} {\bibinfo {author} {\bibfnamefont {P.}~\bibnamefont
  {Beiersdorfer}}, \bibinfo {author} {\bibfnamefont {A.~L.}\ \bibnamefont
  {Osterheld}}, \bibinfo {author} {\bibfnamefont {J.}~\bibnamefont {Scofield}},
  \bibinfo {author} {\bibfnamefont {B.}~\bibnamefont {Wargelin}}, \ and\
  \bibinfo {author} {\bibfnamefont {R.~E.}\ \bibnamefont {Marrs}},\ }\href@noop
  {} {\bibfield  {journal} {\bibinfo  {journal} {Phys. Rev. Lett.}\ }\textbf
  {\bibinfo {volume} {67}},\ \bibinfo {pages} {2272} (\bibinfo {year}
  {1991})}\BibitemShut {NoStop}%
\bibitem [{\citenamefont {L\'opez-Urrutia}\ \emph {et~al.}(2002)\citenamefont
  {L\'opez-Urrutia}, \citenamefont {Beiersdorfer}, \citenamefont {Widmann},\
  and\ \citenamefont {Decaux}}]{Urrutia7}%
  \BibitemOpen
  \bibfield  {author} {\bibinfo {author} {\bibfnamefont {J.~C.}\ \bibnamefont
  {L\'opez-Urrutia}}, \bibinfo {author} {\bibfnamefont {P.}~\bibnamefont
  {Beiersdorfer}}, \bibinfo {author} {\bibfnamefont {K.}~\bibnamefont
  {Widmann}}, \ and\ \bibinfo {author} {\bibfnamefont {V.}~\bibnamefont
  {Decaux}},\ }\href@noop {} {\bibfield  {journal} {\bibinfo  {journal}
  {Canadian Journal of Physics}\ }\textbf {\bibinfo {volume} {80}},\ \bibinfo
  {pages} {1687} (\bibinfo {year} {2002})}\BibitemShut {NoStop}%
\bibitem [{\citenamefont {Windberger}\ \emph {et~al.}(2016)\citenamefont
  {Windberger}, \citenamefont {Torretti}, \citenamefont {Borschevsky},
  \citenamefont {Ryabtsev}, \citenamefont {Dobrodey}, \citenamefont {Bekker},
  \citenamefont {Eliav}, \citenamefont {Kaldor}, \citenamefont {Ubachs},
  \citenamefont {Hoekstra}, \citenamefont {Crespo L\'opez-Urrutia},\ and\
  \citenamefont {Versolato}}]{Windberger2}%
  \BibitemOpen
  \bibfield  {author} {\bibinfo {author} {\bibfnamefont {A.}~\bibnamefont
  {Windberger}}, \bibinfo {author} {\bibfnamefont {F.}~\bibnamefont
  {Torretti}}, \bibinfo {author} {\bibfnamefont {A.}~\bibnamefont
  {Borschevsky}}, \bibinfo {author} {\bibfnamefont {A.}~\bibnamefont
  {Ryabtsev}}, \bibinfo {author} {\bibfnamefont {S.}~\bibnamefont {Dobrodey}},
  \bibinfo {author} {\bibfnamefont {H.}~\bibnamefont {Bekker}}, \bibinfo
  {author} {\bibfnamefont {E.}~\bibnamefont {Eliav}}, \bibinfo {author}
  {\bibfnamefont {U.}~\bibnamefont {Kaldor}}, \bibinfo {author} {\bibfnamefont
  {W.}~\bibnamefont {Ubachs}}, \bibinfo {author} {\bibfnamefont
  {R.}~\bibnamefont {Hoekstra}}, \bibinfo {author} {\bibfnamefont {J.~R.}\
  \bibnamefont {Crespo L\'opez-Urrutia}}, \ and\ \bibinfo {author}
  {\bibfnamefont {O.~O.}\ \bibnamefont {Versolato}},\ }\href@noop {} {\bibfield
   {journal} {\bibinfo  {journal} {Phys. Rev. A}\ }\textbf {\bibinfo {volume}
  {94}},\ \bibinfo {pages} {012506} (\bibinfo {year} {2016})}\BibitemShut
  {NoStop}%
\bibitem [{\citenamefont {Murata}\ \emph {et~al.}(2017)\citenamefont {Murata},
  \citenamefont {Nakajima}, \citenamefont {Safronova}, \citenamefont
  {Safronova},\ and\ \citenamefont {Nakamura}}]{Murata1}%
  \BibitemOpen
  \bibfield  {author} {\bibinfo {author} {\bibfnamefont {S.}~\bibnamefont
  {Murata}}, \bibinfo {author} {\bibfnamefont {T.}~\bibnamefont {Nakajima}},
  \bibinfo {author} {\bibfnamefont {M.~S.}\ \bibnamefont {Safronova}}, \bibinfo
  {author} {\bibfnamefont {U.~I.}\ \bibnamefont {Safronova}}, \ and\ \bibinfo
  {author} {\bibfnamefont {N.}~\bibnamefont {Nakamura}},\ }\href@noop {}
  {\bibfield  {journal} {\bibinfo  {journal} {Phys. Rev. A}\ }\textbf {\bibinfo
  {volume} {96}},\ \bibinfo {pages} {062506} (\bibinfo {year}
  {2017})}\BibitemShut {NoStop}%
\bibitem [{\citenamefont {Tr\"abert}\ \emph {et~al.}(2006)\citenamefont
  {Tr\"abert}, \citenamefont {Beiersdorfer}, \citenamefont {Brown},
  \citenamefont {Boyce}, \citenamefont {Kelley}, \citenamefont {Kilbourne},
  \citenamefont {Porter},\ and\ \citenamefont {Szymkowiak}}]{Trabert7}%
  \BibitemOpen
  \bibfield  {author} {\bibinfo {author} {\bibfnamefont {E.}~\bibnamefont
  {Tr\"abert}}, \bibinfo {author} {\bibfnamefont {P.}~\bibnamefont
  {Beiersdorfer}}, \bibinfo {author} {\bibfnamefont {G.~V.}\ \bibnamefont
  {Brown}}, \bibinfo {author} {\bibfnamefont {K.}~\bibnamefont {Boyce}},
  \bibinfo {author} {\bibfnamefont {R.~L.}\ \bibnamefont {Kelley}}, \bibinfo
  {author} {\bibfnamefont {C.~A.}\ \bibnamefont {Kilbourne}}, \bibinfo {author}
  {\bibfnamefont {F.~S.}\ \bibnamefont {Porter}}, \ and\ \bibinfo {author}
  {\bibfnamefont {A.}~\bibnamefont {Szymkowiak}},\ }\href@noop {} {\bibfield
  {journal} {\bibinfo  {journal} {Phys. Rev. A}\ }\textbf {\bibinfo {volume}
  {73}},\ \bibinfo {pages} {022508} (\bibinfo {year} {2006})}\BibitemShut
  {NoStop}%
\bibitem [{\citenamefont {Roberts}\ \emph {et~al.}(1997)\citenamefont
  {Roberts}, \citenamefont {Taylor}, \citenamefont {Barwood}, \citenamefont
  {Gill}, \citenamefont {Klein},\ and\ \citenamefont {Rowley}}]{Roberts1}%
  \BibitemOpen
  \bibfield  {author} {\bibinfo {author} {\bibfnamefont {M.}~\bibnamefont
  {Roberts}}, \bibinfo {author} {\bibfnamefont {P.}~\bibnamefont {Taylor}},
  \bibinfo {author} {\bibfnamefont {G.~P.}\ \bibnamefont {Barwood}}, \bibinfo
  {author} {\bibfnamefont {P.}~\bibnamefont {Gill}}, \bibinfo {author}
  {\bibfnamefont {H.~A.}\ \bibnamefont {Klein}}, \ and\ \bibinfo {author}
  {\bibfnamefont {W.~R.~C.}\ \bibnamefont {Rowley}},\ }\href@noop {} {\bibfield
   {journal} {\bibinfo  {journal} {Phys. Rev. Lett.}\ }\textbf {\bibinfo
  {volume} {78}},\ \bibinfo {pages} {1876} (\bibinfo {year}
  {1997})}\BibitemShut {NoStop}%
\bibitem [{\citenamefont {Roberts}\ \emph {et~al.}(2000)\citenamefont
  {Roberts}, \citenamefont {Taylor}, \citenamefont {Barwood}, \citenamefont
  {Rowley},\ and\ \citenamefont {Gill}}]{Roberts2}%
  \BibitemOpen
  \bibfield  {author} {\bibinfo {author} {\bibfnamefont {M.}~\bibnamefont
  {Roberts}}, \bibinfo {author} {\bibfnamefont {P.}~\bibnamefont {Taylor}},
  \bibinfo {author} {\bibfnamefont {G.~P.}\ \bibnamefont {Barwood}}, \bibinfo
  {author} {\bibfnamefont {W.~R.~C.}\ \bibnamefont {Rowley}}, \ and\ \bibinfo
  {author} {\bibfnamefont {P.}~\bibnamefont {Gill}},\ }\href@noop {} {\bibfield
   {journal} {\bibinfo  {journal} {Phys. Rev. A}\ }\textbf {\bibinfo {volume}
  {62}},\ \bibinfo {pages} {020501} (\bibinfo {year} {2000})}\BibitemShut
  {NoStop}%
\bibitem [{\citenamefont {Hosaka}\ \emph {et~al.}(2009)\citenamefont {Hosaka},
  \citenamefont {Webster}, \citenamefont {Stannard}, \citenamefont {Walton},
  \citenamefont {Margolis},\ and\ \citenamefont {Gill}}]{Hosaka2}%
  \BibitemOpen
  \bibfield  {author} {\bibinfo {author} {\bibfnamefont {K.}~\bibnamefont
  {Hosaka}}, \bibinfo {author} {\bibfnamefont {S.~A.}\ \bibnamefont {Webster}},
  \bibinfo {author} {\bibfnamefont {A.}~\bibnamefont {Stannard}}, \bibinfo
  {author} {\bibfnamefont {B.~R.}\ \bibnamefont {Walton}}, \bibinfo {author}
  {\bibfnamefont {H.~S.}\ \bibnamefont {Margolis}}, \ and\ \bibinfo {author}
  {\bibfnamefont {P.}~\bibnamefont {Gill}},\ }\href@noop {} {\bibfield
  {journal} {\bibinfo  {journal} {Phys. Rev. A}\ }\textbf {\bibinfo {volume}
  {79}},\ \bibinfo {pages} {033403} (\bibinfo {year} {2009})}\BibitemShut
  {NoStop}%
\bibitem [{\citenamefont {Nakamura}\ \emph {et~al.}(2008)\citenamefont
  {Nakamura}, \citenamefont {Kikuchi}, \citenamefont {Sakaue},\ and\
  \citenamefont {Watanabe}}]{cobit}%
  \BibitemOpen
  \bibfield  {author} {\bibinfo {author} {\bibfnamefont {N.}~\bibnamefont
  {Nakamura}}, \bibinfo {author} {\bibfnamefont {H.}~\bibnamefont {Kikuchi}},
  \bibinfo {author} {\bibfnamefont {H.~A.}\ \bibnamefont {Sakaue}}, \ and\
  \bibinfo {author} {\bibfnamefont {T.}~\bibnamefont {Watanabe}},\ }\href@noop
  {} {\bibfield  {journal} {\bibinfo  {journal} {Rev. Sci. Instrum.}\ }\textbf
  {\bibinfo {volume} {79}},\ \bibinfo {pages} {063104} (\bibinfo {year}
  {2008})}\BibitemShut {NoStop}%
\bibitem [{\citenamefont {Kramida}\ \emph
  {et~al.}(2015{\natexlab{a}})\citenamefont {Kramida}, \citenamefont
  {Ralchenko}, \citenamefont {Reader},\ and\ \citenamefont {{NIST ASD
  Team}}}]{NISTdatabase2015}%
  \BibitemOpen
  \bibfield  {author} {\bibinfo {author} {\bibfnamefont {A.}~\bibnamefont
  {Kramida}}, \bibinfo {author} {\bibfnamefont {Y.}~\bibnamefont {Ralchenko}},
  \bibinfo {author} {\bibfnamefont {J.}~\bibnamefont {Reader}}, \ and\ \bibinfo
  {author} {\bibnamefont {{NIST ASD Team}}},\ }\href@noop {} {\enquote
  {\bibinfo {title} {Nist atomic spectra database (ve{Rev. Sci. Instrum.}on
  5.2) [online]},}\ }\bibinfo {howpublished} {http://physics.nist.gov/asd}
  (\bibinfo {year} {2015}{\natexlab{a}})\BibitemShut {NoStop}%
\bibitem [{\citenamefont {Bar-Shalom}\ \emph {et~al.}(2001)\citenamefont
  {Bar-Shalom}, \citenamefont {Klapisch},\ and\ \citenamefont
  {Oreg}}]{HULLAC2}%
  \BibitemOpen
  \bibfield  {author} {\bibinfo {author} {\bibfnamefont {A.}~\bibnamefont
  {Bar-Shalom}}, \bibinfo {author} {\bibfnamefont {M.}~\bibnamefont
  {Klapisch}}, \ and\ \bibinfo {author} {\bibfnamefont {J.}~\bibnamefont
  {Oreg}},\ }\href@noop {} {\bibfield  {journal} {\bibinfo  {journal} {J.
  Quant. Spectrosc. Radiat. Trans.}\ }\textbf {\bibinfo {volume} {71}},\
  \bibinfo {pages} {169} (\bibinfo {year} {2001})}\BibitemShut {NoStop}%
\bibitem [{\citenamefont {Kramida}\ \emph
  {et~al.}(2015{\natexlab{b}})\citenamefont {Kramida}, \citenamefont
  {{Yu.~Ralchenko}}, \citenamefont {Reader},\ and\ \citenamefont {{and NIST ASD
  Team}}}]{NIST}%
  \BibitemOpen
  \bibfield  {author} {\bibinfo {author} {\bibfnamefont {A.}~\bibnamefont
  {Kramida}}, \bibinfo {author} {\bibnamefont {{Yu.~Ralchenko}}}, \bibinfo
  {author} {\bibfnamefont {J.}~\bibnamefont {Reader}}, \ and\ \bibinfo {author}
  {\bibnamefont {{and NIST ASD Team}}},\ }\href@noop {} {}\bibinfo
  {howpublished} {{NIST Atomic Spectra Database (ver. 5.3), [Online].
  Available: {\tt{http://physics.nist.gov/asd}} [2017, July 6]. National
  Institute of Standards and Technology, Gaithersburg, MD.}} (\bibinfo {year}
  {2015}{\natexlab{b}})\BibitemShut {NoStop}%
\bibitem [{\citenamefont {Jonauskas}\ \emph {et~al.}(2015)\citenamefont
  {Jonauskas}, \citenamefont {Kynien\.e}, \citenamefont {Rynkun}, \citenamefont
  {Ku\v{c}as}, \citenamefont {Gaigalas}, \citenamefont {Kisielius},
  \citenamefont {\v{S} Masys}, \citenamefont {Merkelis},\ and\ \citenamefont
  {Rad\v{z}i\=ut\.e}}]{Jonauskas2}%
  \BibitemOpen
  \bibfield  {author} {\bibinfo {author} {\bibfnamefont {V.}~\bibnamefont
  {Jonauskas}}, \bibinfo {author} {\bibfnamefont {A.}~\bibnamefont
  {Kynien\.e}}, \bibinfo {author} {\bibfnamefont {P.}~\bibnamefont {Rynkun}},
  \bibinfo {author} {\bibfnamefont {S.}~\bibnamefont {Ku\v{c}as}}, \bibinfo
  {author} {\bibfnamefont {G.}~\bibnamefont {Gaigalas}}, \bibinfo {author}
  {\bibfnamefont {R.}~\bibnamefont {Kisielius}}, \bibinfo {author}
  {\bibnamefont {\v{S} Masys}}, \bibinfo {author} {\bibfnamefont
  {G.}~\bibnamefont {Merkelis}}, \ and\ \bibinfo {author} {\bibfnamefont
  {L.}~\bibnamefont {Rad\v{z}i\=ut\.e}},\ }\href@noop {} {\bibfield  {journal}
  {\bibinfo  {journal} {Journal of Physics B: Atomic, Molecular and Optical
  Physics}\ }\textbf {\bibinfo {volume} {48}},\ \bibinfo {pages} {135003}
  (\bibinfo {year} {2015})}\BibitemShut {NoStop}%
\bibitem [{\citenamefont {Safronova}\ and\ \citenamefont
  {Safronova}(2010)}]{Safronova7}%
  \BibitemOpen
  \bibfield  {author} {\bibinfo {author} {\bibfnamefont {U.~I.}\ \bibnamefont
  {Safronova}}\ and\ \bibinfo {author} {\bibfnamefont {A.~S.}\ \bibnamefont
  {Safronova}},\ }\href@noop {} {\bibfield  {journal} {\bibinfo  {journal}
  {Journal of Physics B: Atomic, Molecular and Optical Physics}\ }\textbf
  {\bibinfo {volume} {43}},\ \bibinfo {pages} {074026} (\bibinfo {year}
  {2010})}\BibitemShut {NoStop}%
\bibitem [{\citenamefont {Ding}\ \emph {et~al.}(2012)\citenamefont {Ding},
  \citenamefont {Koike}, \citenamefont {Murakami}, \citenamefont {Kato},
  \citenamefont {Sakaue}, \citenamefont {Dong},\ and\ \citenamefont
  {Nakamura}}]{Ding2}%
  \BibitemOpen
  \bibfield  {author} {\bibinfo {author} {\bibfnamefont {X.-B.}\ \bibnamefont
  {Ding}}, \bibinfo {author} {\bibfnamefont {F.}~\bibnamefont {Koike}},
  \bibinfo {author} {\bibfnamefont {I.}~\bibnamefont {Murakami}}, \bibinfo
  {author} {\bibfnamefont {D.}~\bibnamefont {Kato}}, \bibinfo {author}
  {\bibfnamefont {H.~A.}\ \bibnamefont {Sakaue}}, \bibinfo {author}
  {\bibfnamefont {C.-Z.}\ \bibnamefont {Dong}}, \ and\ \bibinfo {author}
  {\bibfnamefont {N.}~\bibnamefont {Nakamura}},\ }\href@noop {} {\bibfield
  {journal} {\bibinfo  {journal} {J. Phys. B}\ }\textbf {\bibinfo {volume}
  {45}},\ \bibinfo {pages} {035003} (\bibinfo {year} {2012})}\BibitemShut
  {NoStop}%
\bibitem [{\citenamefont {Safronova}\ \emph {et~al.}(2003)\citenamefont
  {Safronova}, \citenamefont {Savukov}, \citenamefont {Safronova},\ and\
  \citenamefont {Johnson}}]{Safronova6}%
  \BibitemOpen
  \bibfield  {author} {\bibinfo {author} {\bibfnamefont {U.~I.}\ \bibnamefont
  {Safronova}}, \bibinfo {author} {\bibfnamefont {I.~M.}\ \bibnamefont
  {Savukov}}, \bibinfo {author} {\bibfnamefont {M.~S.}\ \bibnamefont
  {Safronova}}, \ and\ \bibinfo {author} {\bibfnamefont {W.~R.}\ \bibnamefont
  {Johnson}},\ }\href@noop {} {\bibfield  {journal} {\bibinfo  {journal} {Phys.
  Rev. A}\ }\textbf {\bibinfo {volume} {68}},\ \bibinfo {pages} {062505}
  (\bibinfo {year} {2003})}\BibitemShut {NoStop}%
\end{thebibliography}%

\end{document}